\documentclass[prl,superscriptaddress,twocolumn,showpacs]{revtex4}
\usepackage{epsfig,amsmath,amssymb,graphics,color,calc}

\newcommand{\be}{\begin{equation}}
\newcommand{\ee}{\end{equation}}
\newcommand{\ba}{\begin{eqnarray}}
\newcommand{\ea}{\end{eqnarray}}

\begin{document}

\title{Universal nature of particle displacements 
close to glass and jamming transitions}

\author{Pinaki Chaudhuri}
\affiliation{Laboratoire des Collo{\"\i}des, Verres
et Nanomat{\'e}riaux, UMR 5587, Universit{\'e} Montpellier II and CNRS,
34095 Montpellier, France}

\author{Ludovic Berthier}

\affiliation{Laboratoire des Collo{\"\i}des, Verres
et Nanomat{\'e}riaux, UMR 5587, Universit{\'e} Montpellier II and CNRS,
34095 Montpellier, France}

\affiliation{Joint Theory Institute, Argonne National Laboratory and
University of Chicago,
5640 S. Ellis Av., Chicago, Il 60637}

\author{Walter Kob}
\affiliation{Laboratoire des Collo{\"\i}des, Verres
et Nanomat{\'e}riaux, UMR 5587, Universit{\'e} Montpellier II and CNRS,
34095 Montpellier, France}

\date{\today}

\begin{abstract}
We examine the structure of the distribution of single particle displacements 
(van-Hove function) in a broad class of materials close to
glass and jamming transitions. In a wide time window comprising 
structural relaxation, van-Hove functions reflect the coexistence 
of slow and fast particles (dynamic heterogeneity). 
The tails of the distributions exhibit exponential, rather than Gaussian, 
decay. We argue that this behavior is universal in 
glassy materials and should be considered the analog, in space, 
of the stretched exponential decay of time correlation functions.  
We introduce a dynamical model that describes quantitatively 
numerical and experimental data in supercooled liquids, 
colloidal hard spheres and granular materials. The tails of the distributions
directly explain the decoupling between translational 
diffusion and structural relaxation observed in glassy materials. 
\end{abstract}





\maketitle

The slow dynamics of disordered materials
close to glass and jamming transitions is characterized 
by just a few universal features~\cite{review}: dramatic 
dynamical changes upon mild changes of control parameters
(temperature, density), broad distribution 
of relaxation times leading to  
stretched exponential decay of time correlation functions,
and spatially heterogeneous dynamics~\cite{ediger}.
Here we argue that the detailed structure
of the distribution of particles displacements (van-Hove
function~\cite{vanhove}) 
constitutes an additional universal signature 
of glassy dynamics. We show that, for timescales
corresponding to structural relaxation, 
the self part of the van-Hove function 
has fat tails that are well-described by an exponential, 
rather than a Gaussian, decay. We provide a broad range
of numerical and experimental data, physical
arguments, and a dynamical model to support this claim.

The non-Fickian character of single particle displacements 
in materials with glassy dynamics is 
well-known~\cite{rahman,odagaki,kob,weeks,marty,virgile,laura,pablo,stariolo}:  
time correlation functions decay non-exponentially, mean-squared displacements
exhibit at intermediate timescales a subdiffusive plateau, 
van-Hove distributions  are non-Gaussian. This affects transport
properties since translational diffusion is decoupled from structural 
relaxation~\cite{decouplingexp}, 
leading to an anomalous relation between timescales
and lengthscales~\cite{berthier}.
Virtually all glass theories address the stretched decay 
of correlation functions, but comparatively 
much less attention has been paid to the detailed shape
of the self-part of the 
van-Hove function~\cite{odagaki,archer,schweizer,epl,langer,stariolo}, 
although new techniques now directly
access this quantity in different 
materials~\cite{weeks,marty,virgile,laura}. 
Its non-Gaussian, ``heterogeneous'' character
is often discussed in qualitative terms~\cite{archer,schweizer}, 
and quantitative measures focus 
on the distribution kurtosis (non-Gaussian 
parameter~\cite{rahman}) which contains, however, very indirect information 
about its shape.
Deviations from Gaussian behavior are usually ascribed 
to dynamic heterogeneity~\cite{ediger}, i.e. to the presence of particles 
that are substantially faster or slower than the average.  
We argue that van-Hove functions contain quantitatively relevant
information about the relaxation of glassy materials, and 
that its functional form is simple and 
universal, just as the stretched exponential
decay of time correlation functions. Glass theories
should therefore treat both phenomena on an equal footing.

\begin{figure}
\psfig{file=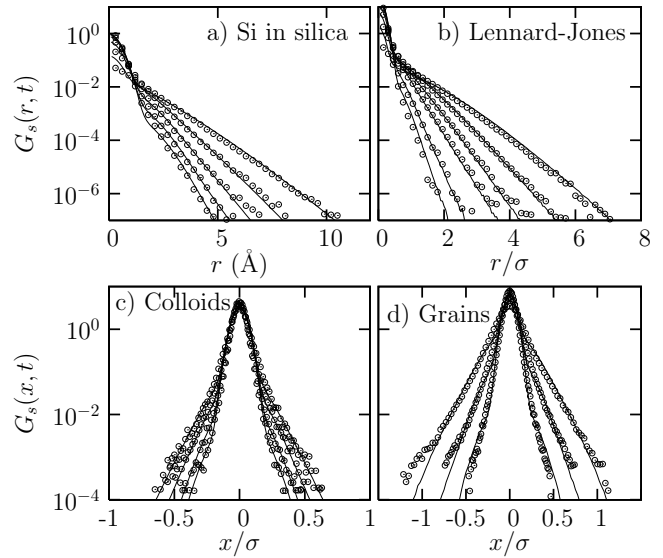,width=8.5cm}
\caption{\label{4panel} Time evolution of the 
self part of van-Hove functions for silicon atoms in silica,
Lennard-Jones particles, hard-sphere colloids and grains 
(open circles), fitted with the model 
in Eq.~(\ref{model}) (full lines). 
They exhibit a Gaussian central part and a fat, exponential tail. 
(a) $T=3000$~K and $t \in [27, 1650]$~ps.
(b) $T=0.435$ and $t \in [75 \cdot 10^3, 41 \cdot 10^6]$.
(c) $\varphi = 0.517$ and $t \in [90, 1008]$~s. 
(d) $\varphi = 0.84$ and $t \in [10,1000]$ cycles.
(a) and (b) show the distributions of $|{\bf r}(t)-{\bf r}(0)|$, 
(c) and (d) the distributions of $x(t)-x(0)$. 
}  
\end{figure}
  
\begin{figure}
\psfig{file=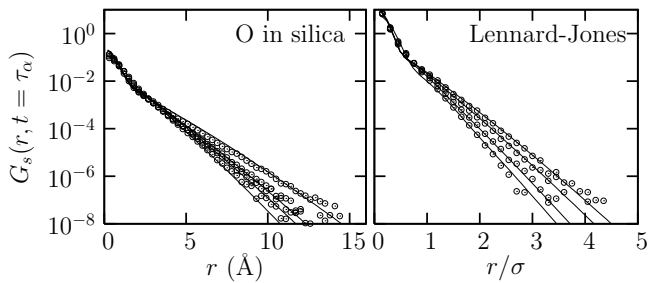,width=8.5cm}
\caption{\label{2panel} Temperature evolution 
of the self part of the van-Hove 
function for $t=\tau_\alpha$ for 
oxygen atoms in silica at $T=3580$, 3200, 3000 and 2715~K 
(from left to right) and Lennard-Jones particles at
$T=0.5$, 0.47, 0.45 and 0.435 (from left to right).
The exponential tail gets more pronounced
at low temperatures. This trend is smaller in silica than for 
Lennard-Jones, as is the translational decoupling.}
\end{figure}

We present our central observations in Figs.~\ref{4panel}, 
\ref{2panel}
which show the self part of the van-Hove functions 
for  a silica melt~\cite{bks},
a binary Lennard-Jones (LJ) mixture~\cite{lj},  a dense suspension 
of colloidal hard spheres~\cite{weeks}, and  a slowly driven dense granular 
assembly~\cite{marty}. It reads~\cite{vanhove} 
\be
G_s (r,t) = \langle  \delta ( r - |{\bf r}_i(t) - {\bf r}_i(0)|) \rangle,
\label{gs}
\ee 
where ${\bf r}_i(t)$ denotes the position of a particle $i$ (molecule, 
colloid, grain) at time $t$, the brackets indicate an ensemble
average. For technical reasons, experiments sometimes 
record the one-dimensional version of (\ref{gs}), 
$G_s(x,t) = \langle  \delta ( x - |x_i(t) - x_i(0)|) \rangle$,
where $x_i(t)$ is the projection of ${\bf r}_i(t)$ on a given unit vector.
For our purposes,
the difference between the two functions is irrelevant.
For all systems we find that $G_s(r,t)$ has the same structure
over a broad time window comprising the structural relaxation.
Most of its statistical 
weight is carried by particles that have barely moved, 
$r < \sigma$, but a ``fat'' tail extends to much 
larger distances,
$r > \sigma$, where $\sigma$ is the particle diameter. 
The small $r$ behavior is not far from a Gaussian
distribution, corresponding to quasi-harmonic vibrations,
but the large distance decay 
appears linear in Fig.~\ref{4panel}, i.e. 
$G_s \propto \exp(-r/\lambda(t))$. For a Fickian particle, 
one expects instead a Gaussian decay, 
$G_s \propto \exp(-r^2/ (4 D_s t))$, where 
$D_s$ is the self-diffusion constant. 
Although the existence of a fat tail in $G_s(r,t)$ was 
recognized 
before~\cite{weeks,marty,virgile,laura,kob,archer,schweizer,stariolo,epl,szamel1}, 
its non-trivial functional form and universality
went largely unnoticed.

The exponential tail extends to larger distances when $t$ increases, but
$\lambda(t)$ grows very slowly with $t$. 
Eventually, at very large times, a crossover to Fickian
behavior is observed, see the latest time in Fig.~\ref{4panel} b.
For the LJ system the crossover takes place 
for $t \approx 30 \tau_\alpha$~\cite{szamel1}, where $\tau_\alpha(T)$
is defined from the time decay of the self-intermediate scattering
function, $F_s(q,t)$, the Fourier transform of $G_s(r,t)$. 
Our observations correspond to times that are shorter 
than this crossover.
In Fig.~\ref{2panel}, we present the evolution of $G_s(r,t)$ 
when the glass transition is approached, keeping $t$ 
fixed to $\tau_\alpha(T)$. 
Clearly the shape of $G_s(r,t)$ remains unchanged, but 
the tail becomes more pronounced closer to the glass transition.

\begin{figure}
\psfig{file=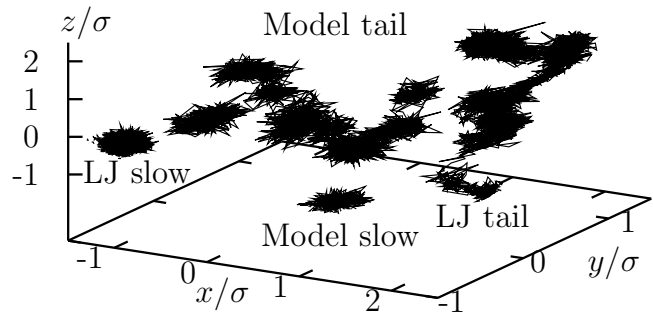,width=8.5cm}
\caption{\label{snapshot} Four trajectories 
of duration $\tau_\alpha$
for particles contributing to the center of the distribution (``slow'')
and to the tail (``tail'') taken from simulations of the 
LJ liquid at $T=0.435$ (``LJ'')
and for random walkers described by (\ref{model}).}
\end{figure}

These observations confirm that  van-Hove 
functions can be qualitatively  described as the superposition of 
two families of particles: localized particles 
contributing to the Gaussian central part and
mobile particles contributing to the exponential tail. 
Evidence has recently been given that
the distinction between mobile and immobile particles cannot 
be explained on a structural basis~\cite{rob}. We 
therefore seek a purely dynamical explanation.
In Fig.~\ref{snapshot} we present representative trajectories
of duration $\tau_\alpha$ for mobile and immobile particles in the LJ system.
Similar pictures have been presented before~\cite{weeks,marty}. 
A large fraction of the particles perform 
localized, vibrational motion around their initial positions,
as in a disordered solid. These 
``slow'' particles are ``caged'' by their neighbors 
and contribute to the quasi-Gaussian 
central part of $G_s(r,t)$. 

More interesting is the behavior
of the particles contributing to the tail. On top of the 
localized vibrations, these particles perform a 
(distributed) number of quasi-instantaneous 
``jumps''. 
This suggests that particles 
perform a form of continuous time 
random walk (CTRW~\cite{mw})~\cite{elliott,odagaki,epl,langer}. 
From direct inspection of the trajectories we note that
the size of the jumps is distributed, and represents
on average only a small fraction of the particle size, 
implying that jumps probably result from cooperative events involving
a large number of particles moving by a small amount~\cite{demos}. 
Regarding timescales, trajectories also reveal that 
the time of the first jump after $t=0$ is distributed. This 
observation  directly implies
that van-Hove functions can be described, for $t>0$, 
as a superposition between particles that have jumped, and those that 
have not~\cite{epl}. We insist~\cite{pablo,rob} that this coexistence 
is dynamically generated and we will avoid the assumption 
of a material being composed of two dynamically 
distinct families of particles~\cite{2fluid,langer}.

The final empirical observation from Fig.~\ref{snapshot}
is that once a particle has managed to make a jump, 
it very likely makes one or several additional jumps 
during the rest of our observation time.
We believe that this results from dynamic heterogeneity.
Spatial clustering of particles with correlated dynamics implies indeed
that it takes a very long time for a particle belonging to a slow
region to become mobile. But when this happens, 
the particle then likely belongs to 
a mobile region, which enhances considerably its probability 
to move further. Different timescales for initial 
and subsequent moves is an exact result, for the reason mentioned above,
in kinetically constrained models~\cite{jung,epl}. It is likely a 
generic consequence of the presence of spatially 
heterogeneous dynamics. 

These features endow particle diffusion 
in glassy materials with specific properties. 
We now introduce a model which incorporates
these empirical observations with as few free parameters 
as possible. We significantly extend the work of Ref.~\cite{epl}, which 
analyzed self-diffusion in kinetically constrained models,
to describe off-lattice realistic models and experiments.
The system is viewed as an assembly of 
dynamically indistinguishable particles, 
compatible with structural homogeneity.
We assume solid behavior at short times.
In between jumps, particles perform
on a microscopic timescale a Gaussian exploration 
of their environment with the distribution 
$f_{\rm vib}(r) = (2 \pi \ell^2)^{-3/2} 
\exp(-r^2/2 \ell^2)$. We assume that particles 
perform jumps with a size sampled from
$f_{\rm jump}(r) = (2 \pi d^2)^{-3/2}
\exp(-r^2/2 d^2)$. We similarly assume simple forms 
for time distributions. The time of the first jump
is drawn from $\phi_1 (t) = \tau_1^{-1} \exp(-t/\tau_1)$. We then assume that
subsequent jumps arise with higher frequency, 
using the distribution $\phi_2 (t) = \tau_2^{-1} \exp(-t/\tau_2)$, 
with $\tau_2 < \tau_1$. 
It is now a simple
task~\cite{mw} to express the van-Hove function,
$G_s(r,t) = \sum_{n=0}^{\infty} p(n,t) f(n,r)$, where
$p(n,t)$ is the probability to make $n$ jumps 
in a time $t$, and $f(n,r)$ the probability to move a distance 
$r$ in $n$ jumps. These probabilities involve convolutions 
and are more easily expressed in the Fourier-Laplace domain, 
($r,t$) $\to$ ($q,s$). We obtain
\be
G_s(q,s) = f_{\rm vib}(q) \Phi_1(s)    + 
f(q) f_{\rm vib}(q)  \frac{ \phi_1(s)  \Phi_2(s)}{1-\phi_2(s)
f(q)},  
\label{model}
\ee 
where 
$\phi_{1,2}(s) \equiv 1- s \Phi_{1,2}(s)$ and
$f(q) \equiv f_{\rm vib}(q) f_{\rm jump}(q)$. 
The Montroll-Weiss 
equation~\cite{mw} is recovered when $\phi_1=\phi_2$ 
and vibrations are not considered, $f_{\rm vib}(q)=1$.
The result in Eq.~(\ref{model}) is valid for 
any choice of distributions ($f_{\rm vib}$, 
$f_{\rm jump}$, $\phi_1$, 
$\phi_2$). Here, we restrict to simple choices (Gaussian and exponential
distributions) to emphasize the universality 
and physical origin of our results and to introduce
as few free fitting parameters as possible: 
($\ell$, $d$, $\tau_1$, $\tau_2$). Equation (\ref{model})
makes very transparent the fact that $G_s$
is the superposition of localized particles,  
and mobile particles. 
We show below that the second term produces 
a tail that is close to exponential and arises from 
particles which have performed one or several jumps. 

For the four systems considered in Fig.~\ref{4panel} we have 
used Eq.~(\ref{model}) to fit the self part of the van-Hove functions, as  
shown with full lines. The fits evidently 
match the data very well. In practice, we sought 
the set of parameters that allows for data fitting on the largest 
time window comprising structural relaxation. We find that 
fitting several times fixes the set of parameters
with little ambiguity, while multiple choices remain possible 
when fitting data for a single $t$. 
In Fig.~\ref{snapshot} we present 
numerically generated trajectories
of the generalized CTRW model of (\ref{model}) using the parameters
used to fit the LJ data, leaving us with no doubt that such a model 
captures the main qualitative aspects of the real trajectories. 
As expected, 
we find that both cage and jump sizes represent 
only a fraction of the particle size, and are very weakly dependent
on the control parameters.
As a rule of thumb we find 
$d \approx 2 \ell$. For instance ($d/\sigma$, $\ell/\sigma$) 
is (0.1, 0.051) in  colloids, (0.15, 0.06) in grains,  
and (0.35, 0.15) in the Lennard-Jones. Moreover, our results 
for $d^2$ agree well with plateau 
values directly measured in mean-squared displacements for all systems.  

\begin{figure}
\psfig{file=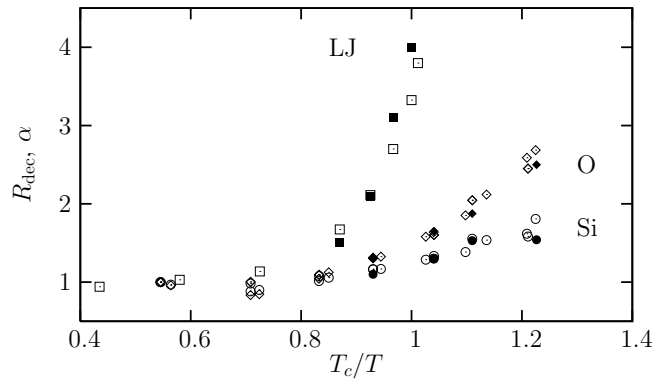,width=8.5cm}
\caption{\label{dec} Comparison of the amount of 
translational decoupling in silica and LJ systems measured in MD simulations
through $R_{\rm dec}$ (open symbols), and of the ratio $\alpha = t_1/t_2$ 
(filled symbols) employed to fit the data with the model (\ref{model}). 
The agreement is excellent.
}
\end{figure}

These results imply that the changes observed in Fig.~\ref{2panel}
are mostly due to a change in the balance between $\tau_1$ and $\tau_2$. 
We naturally find that, in a first approximation, 
both times track the structural relaxation 
of the system. More interesting is the evolution of their 
ratio $\alpha=\tau_1/\tau_2$, 
reported in Fig.~\ref{dec}.
Unfortunately, we had not enough
data for grains and colloids to report
accurate estimates of $\alpha$ in these systems and thus we concentrate
on the two numerical models.
In order to account for the increasingly fat tails reported in 
Fig.~\ref{2panel}, $\alpha$ has to grow 
significantly when temperature decreases, as expected 
from the above  discussion. 
The growth of $\alpha$ directly impacts 
on transport properties~\cite{epl,jung}.
Within the model (\ref{model}) the time 
decay of $F_s(q,t)$ at large $q$ (small distance) 
is governed by $\phi_1(t)$, so that $\tau_\alpha \sim \tau_1$.  
Fickian diffusion is recovered when the average number of jumps
becomes large. It is easy to show 
from the $q, s \to 0$ limit of 
(\ref{model}) that $D_s \sim (\ell^2 + d^2) / \tau_2$, so that
the product between self-diffusion constant and structural relaxation 
time scales as 
$D_s \tau_\alpha \sim \alpha$. Our model  
therefore makes a direct prediction 
about translational decoupling.  We have measured 
the normalized product, $R_{\rm dec} (T) = 
D_s(T) \tau_\alpha(T) /D_s(T_0) \tau_\alpha(T_0)$ (where 
$T_0$ is a high temperature) 
for Lennard-Jones, silicon and oxygen atoms 
directly in numerical simulations, see Fig.~\ref{dec}. 
This ratio is $R_{\rm dec} = 1$ 
at high temperatures, and becomes 
$R_{\rm dec} >1$ whenever 
translational decoupling occurs~\cite{berthier}. 
For three types of particles with different degrees of decoupling, we 
find quantitative agreement between $\alpha$ 
obtained from fits of the self part of the van-Hove function
and the decoupling $R_{\rm dec}$
directly measured in the simulations. 
Thus, Fig.~\ref{dec} gives strong support to our 
physical description and empirical modeling 
of self-diffusion close to a glass transition, and provides 
a quantitative link between dynamic heterogeneity and decoupling.

Why are the tails of the distributions 
described by an exponential decay? Non-Gaussian decay is in fact 
present in the original CTRW model when distances outside the realm
of central limit theorem are considered. These tails
are enhanced, and hence more easily detectable, 
when $\alpha > 1$ and decoupling occurs. 
Consider the case $\alpha=1$, $\ell = 0$
in Eq.~(\ref{model}). Inverting the Laplace transform
yields
\be
G_s(r,t) = G_0 + 
\frac{4 \pi e^{-\bar{t} }}{r} \int_0^\infty dq  
[ e^{ \bar{t}f(q)}  - 1]  q \sin(qr),
\label{integral}
\ee
where 
$G_0(r,t) \equiv \delta(r) \Phi_1(t)$ and $\bar{t} \equiv t/\tau_1$.
We then expand the exponential in (\ref{integral}), 
integrate each term and convert the sum into an integral to get 
\be
G_s(r,t) = G_0(r,t) 
+ \frac{\pi e^{-\bar{t}}  }{4 d^3}  \int_1^\infty dn \frac{e^{-f(n)}}{
n^2} ,
\label{integral2}
\ee
with $f(n) = n \ln n -n \ln \bar{t} -n + 
r^2/(8 d^2 n)$. The large distance limit 
of (\ref{integral2}) is evaluated using a
saddle point approximation, 
\be
G_s(r,t) \sim  
\frac{ (\pi Y)^{3/2}  e^{-\bar{t}} }{ 
(rd)^{3/2}  \sqrt{1+Y^2}  } 
e^{- r [ Y - 1/Y ]/2d},   
\label{tail}
\ee 
where $Y$ satisfies $Y^2 \exp Y^2 = r^2 / (2 d \bar{t})^2$; 
$Y^2 \sim 2 \log (\frac{r}{2d\bar{t}})$ for large $r$. 
Thus, we find that $G_s(r,t)$ decays exponentially 
(with logarithmic corrections) at large $r$.
Interestingly this expansion can be obtained 
independently of the actual shape of the distributions,
establishing its universality. 
Considering that the tail of $G_s$ stems 
from particles that have performed a number
of jumps larger than average, one finds 
$p(n,t) \sim  (\bar{t}/n)^n$  and
$f(n,r) \sim e^{-r^2/(8 d^2 n)}$, yielding 
an expression similar to (\ref{tail}). 

We have reported the existence of a new universal feature characterizing the 
dynamics of materials close to glass and jamming transitions, seen
in the structure of the distribution of single particle displacements
which exhibits exponential decay at large distances. 
We argued it is a generic consequence
of the existence of spatially heterogeneous dynamics, which has
profound consequences on transport properties.
Our results apply to a wide variety of materials 
from atomic liquids~\cite{review} to biophysical materials~\cite{virgile} 
and grains~\cite{marty}. We strongly encourage more 
systematic experimental exploration  
of particle displacements in amorphous materials
with slow dynamics.

\begin{acknowledgments}
We thank O. Dauchot, G. Marty, and E. Weeks for providing 
their data, J.-P. Bouchaud, P. Mayer, 
D. Reichman and G. Szamel for useful discussions.
Financial support from the Joint Theory Institute 
(Argonne National Laboratory and University of Chicago), 
CEFIPRA Project 3004-1, and ANR Grant TSANET is acknowledged. 

\end{acknowledgments}

\end{document}